\documentclass[amsmath,amssymb,prb]{revtex4}
\usepackage{graphicx}
\usepackage{dcolumn}
\usepackage{bm}
\usepackage{url}
\def\figno#1{Fig.~\ref{fig:#1}}
\begin{document}
\title{A small tabletop experiment for a direct measurement of 
  the speed of light}
\author{
  Kenichiro Aoki\footnote{E--mail:~{\tt ken@phys-h.keio.ac.jp}.}
  and Takahisa Mitsui\footnote{E--mail:~{\tt mitsui@hc.cc.keio.ac.jp}.}}
\affiliation{Dept. of Physics, Keio University, {\it
    4---1---1} Hiyoshi, Kouhoku--ku, Yokohama 223--8521, Japan}
  \date{\today }

\begin{abstract}
    A small tabletop experiment for a direct measurement of the speed
    of light to an accuracy of few percent is described. The
    experiment is accessible to a wide spectrum of undergraduate
    students, in particular to students not majoring in science or
    engineering.  The experiment may further include a measurement of
    the index of refraction of a sample. Details of the setup and
    equipment are given.  Results and limitations of the experiment
    are analyzed, partly based on our experience in employing the
    experiment in our student laboratories. Safety considerations are also
    discussed.
\end{abstract}

\vspace{3mm}

\maketitle

\section{Introduction}
\label{sec:intro}
The speed of light is a basic physics quantity of importance in many
areas of physics, be it astrophysics or relativity. Therefore, it is
highly desirable to have an experiment to measure the speed of light
that is easy to set up and any student can perform.  In this note, we
provide details of an experiment for directly measuring the speed of
light in air. This small tabletop experiment can be performed quite
satisfactorily by undergraduate students, including non-science or
non-engineering majors and perhaps also by high school students.
The principle of the experiment dates back at least a few hundred
years to that of Galileo.  We obtain the speed of light from the time
of flight, essentially by measuring the time the light takes to travel
a certain distance and back. The experiment utilizes a pulsed laser
and an oscilloscope to measure the time of flight. 

The requirements for our experiment were as follows:
\begin{itemize}
  \item The concept of the experiment is quite comprehensible to
    non-science and non-engineering majors.
  \item The optical path is in air and is visible.
  \item The experiment can be performed on a tabletop.
  \item The students are able to {\it safely} conduct the experiment
    without difficulty.
  \item The measurement is reasonably accurate, but does not sacrifice
    clarity for precision. 
  \item The cost of the experimental setup is inexpensive. 
\end{itemize}
We have been able to satisfy all the requirements above.  To our
knowledge, standard experimental textbooks do not contain expositions
of an experiment such as ours. Many experiments to measure the speed
of light have been presented, including a number of interesting time
of flight
measurements\cite{Cooke68,Tyler69,Van73,Crandall82,Becc87,Deb91,James99}.
However, we could not find any that satisfied all the requirements. In
particular, most of the simpler experiments require a large space,
which we can not afford.

The purpose of this note is pragmatic: Our objective is to provide
the details of the experiment to a wide audience, so that it can be
performed at other institutions, if so desired.  The electronic part
of the experiment is original and we provide its full details below.
The components needed to build the light source and the detector are
quite inexpensive.
Experiments that are different, but similar in spirit, seem to be
performed at other institutions\cite{Sakurai,tohoku,shimadzu}.
\section{Details of the experiment}
\label{sec:exp}
\begin{figure}[htbp]
    \centering    
     \includegraphics[width=12cm,clip]{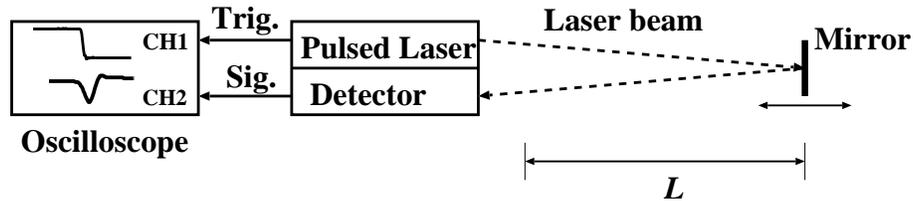}    
    \caption{The schematics of the experiment.}
    \label{fig:setup}
\end{figure}
\begin{figure}[htbp]
    \centering    
     \includegraphics[width=10cm,clip]{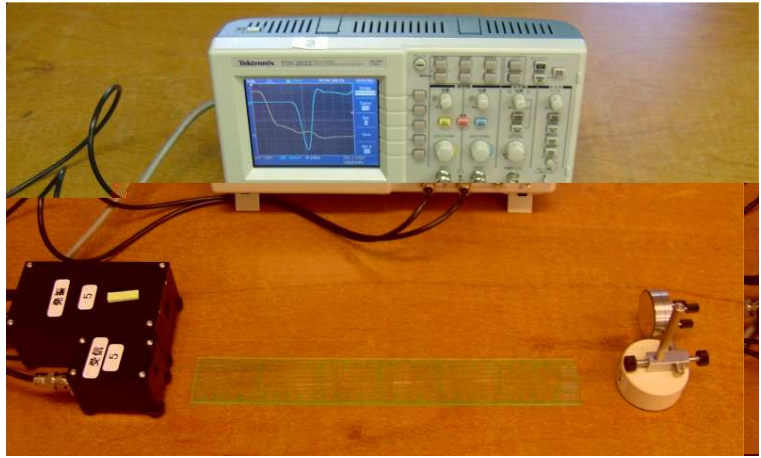}    
    \caption{A setup of the experiment.  The ruler is 32\,cm long and
      is included  here to indicate the scale.}
    \label{fig:setupPhoto}
\end{figure}
\begin{figure}[htbp]
    \centering
     \includegraphics[width=14cm,clip]{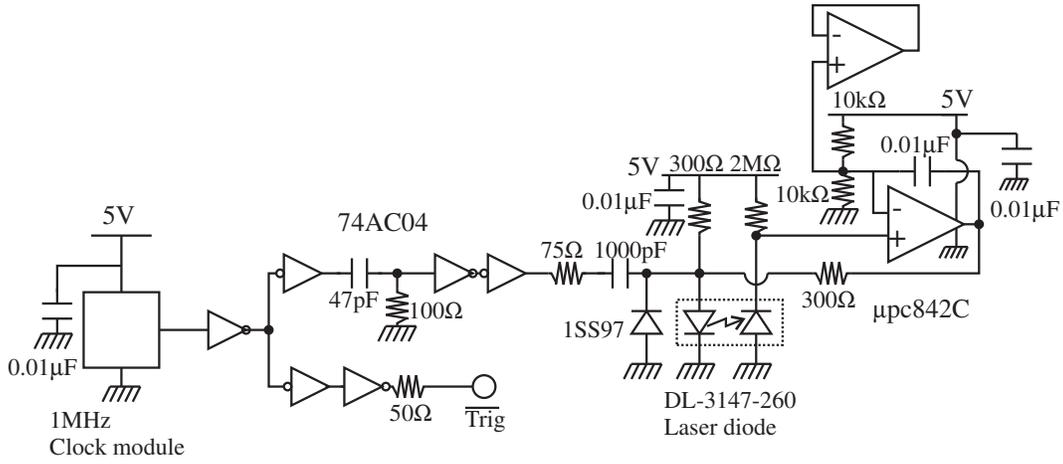}    
    \caption{Circuit diagram for the pulsed laser.}
    \label{fig:laser}
\end{figure}
\begin{figure}[htbp]
    \centering
    \includegraphics[width=8cm,clip]{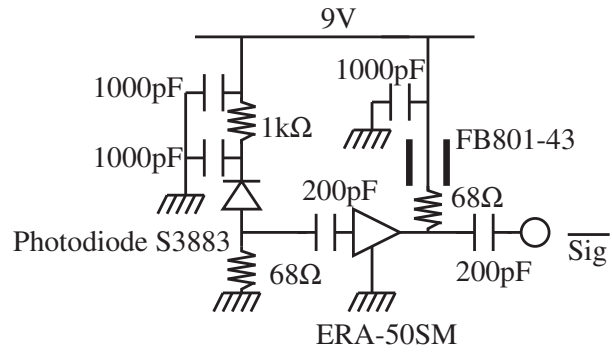}    
    \caption{Circuit diagram for the detector.}
    \label{fig:detector}
\end{figure}
\begin{figure}[htbp]
    \centering
     \includegraphics[width=8.5cm,clip]{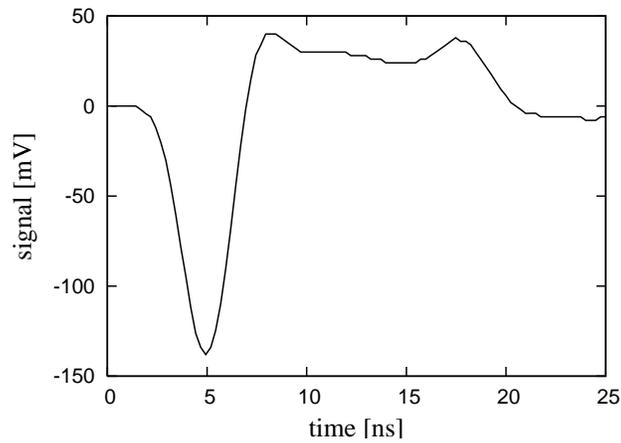}        
    \caption{An example of the detected signal.}
    \label{fig:signal}
\end{figure}

The basic experimental setup is simple and is shown in \figno{setup}
and \figno{setupPhoto}.  Pulsed laser is reflected off a mirror and is
detected by a photodiode. An oscilloscope is used to obtain the time
of flight $\tau=2L/c$, where $L$ is the (one way) length of the path.
The circuits for the pulsed laser and the detector are shown in
\figno{laser} and \figno{detector}.  The pulsed laser has a quartz
generated pulse modulation frequency of 1.0\,MHz, wavelength 650\,nm
and an average optical output of 30\,$\mu$W (30\,pJ/pulse).  We have
used a 200\,MHz oscilloscope with 2\,GHz sampling (Tektronix TDS2022).
Such oscilloscopes are now quite standard and are affordable,
especially considering that they can be used for various purposes.  An
example of the observed signal is shown in \figno{signal}. The total
space required by the experimental equipment is about 1.5\,m by
0.5\,m.

To be concrete, we outline the procedure in our student experiments,
which we have found to be quite effective, as an example.  We let the
students measure the difference in the time of flight, $\tau$, at
various $L$, from a reference value. (We use $L=0.2,0.4,0.6,0.8,1$\,m
from the reference point.)  Then the students plot the data points and
fit a straight line to them to obtain the speed of light.  Some
reasons we have chosen this procedure is as follows:
\begin{itemize}
  \item By measuring the delay caused by light transmission for
    various $L$, it is difficult for students to miss the meaning of
    the finiteness of the speed of light.
  \item Taking multiple measurements not only increases accuracy but
    decreases the probability of simple errors.
\end{itemize}
We remark here that using lengths such as $L=1.5\,$m lead to $\tau_{}$
values which are quite close to round numbers. This tends to mislead
the students into thinking that the speed of light measurement has
unreasonably small experimental errors, since the students tend to
obtain $3.0\times10^8\,$m/s as a result which differs from the precise
value only by 0.1\,\%.

The dominant source of error in the experiment is the measurement of
the time interval, $\tau_{}$. With a 200\,MHz oscilloscope we
use, error in the time interval $\Delta \tau_{}\simeq0.2$\,ns.  With
$L=1$\,m and $\tau_{} =6.7$\,ns, the relative error in the velocity of
light is 3\%.  The error can be reduced by using larger $L$ or a
faster oscilloscope.  Also, using a slower oscilloscope such as one at
100\,MHz increases the experimental error with the same distance.
A systematic error is introduced by approximating the path length by
$L$ without taking into account the lateral shift in the beam. While
the correction can be incorporated, its relative contribution is at a
few times $0.1\,\%$ level.
Overall, we find the precision of this experiment to be quite
satisfactory for our purposes.

We briefly explain an extension of the experiment which we also
perform. By including some refractive material in the optical path, we
can measure its index of refraction from the time delay, $\tau_{\rm
  delay}$, using the equation
\begin{equation}
    \label{eq:refraction}
    c\tau_{\rm delay}=(n-1)L_1
\end{equation}
Here, $L_1$ is the physical length of the refractive material in the
optical path.  We have used an acrylic rod of $L_1=0.75\,$m, which has
an index of refraction, $n=1.49$.  The relative error in $n-1$ is
20\,\% (for $L_1=0.75\,$m), so that the absolute error in the index of
refraction for our setup is $0.1$. This is not so accurate, but is
quite sufficient for our purposes, which is to see that having an
index of refraction larger than one is equivalent to having a slower
speed of light and to obtain the approximate value of the index.
Since the light is attenuated by the refractive material, it is more
difficult to pick up the signal, but students can do so with a little
effort.  This part of the experiment is easier to perform without a
mirror.

Let us comment on the safety aspects regarding the laser beam: The
average power of the laser light is $30\,\mu$W.
The measured pulse width is 2.5\,ns. This is an upper bound since the
high frequency measurement can be limited by the performance of the
oscilloscope itself.  While it is ill--advised to observe the laser
light without safety glasses, the laser beam does not seem to
necessitate such safety precautions at this time\cite{HS04}.  The
laser circuit (\figno{laser}) is designed containing a feedback
circuit to keep the average light output constant. The average output
was chosen to satisfy the following two competing conditions: From
safety considerations, we would like the output to be small. However,
we also would like the beam to be strong enough and visible so that
the experiment can be performed with ease. Our choice seems to satisfy
both these conditions, as explained above. The pulse modulation
frequency was chosen to obtain the desired light output.
\section{Discussion}
\label{sec:disc}
We have explained concretely how to realize a small tabletop experiment
to measure the speed of light to a few percent level.  The error can
be easily brought down to a percent level if we allow ourselves to use
a path length of few meters, instead of one as was done above.  To
achieve such accuracy, the quality of the signal and the detector was
crucial and the required electronics has been explicitly described
above.
We have adopted this experiment, including the index of refraction
measurement, in our classes and roughly 600 students majoring in
humanities and social sciences (mostly freshmen) at Keio University
have performed it. The experiments have been conducted essentially
with no problems and the results are quite satisfactory.  The
experiment is done in groups of one or two students each, in a three
hour session. This includes the time for explanation by the lecturer
and the time to write a short report on the experiment.
The concept of the experiment is easily grasped by
non-science/engineering majors.  Furthermore, the experimental output
tends to lead to a good understanding of the finiteness of speed of
light, which is otherwise not easy to experience directly. Most
students achieve the kind of experimental accuracy as designed.  Also,
using refractive material, the students can really see and understand
that the refractive material ``slows the light down''. For most
non-science/engineering students, it is the first (and maybe the last)
time for them to handle an oscilloscope and this technical aspect is
probably the most difficult part of the experiment.
Most students seem to understand and enjoy the
visualization of the signal provided by the oscilloscope. Other than
some instructions on the usage of the oscilloscope, the experiment can
be performed independently by students in most cases.

Acknowledgements: %
This work was supported in part by the
grants from Keio University and the Ministry of Education, Culture,
Sports, Science and Technology of Japan.


\begin{thebibliography}{99}
  \bibitem{Cooke68}J. Cooke, M. Martin, . McCartney and H. Wilf,
    ``Direct determination of the speed of light as a general physics
    laboratory experiment'', Am. J. Phys. 36, 847 (1968).
  \bibitem{Tyler69}C.E. Tyler, ``A pedagogical Measurement of the
    velocity of light'', Am. J. Phys. 49, 740--745 (1969).
  \bibitem{Van73}J. Vanderkooy and M.J. Beccario, ``An inexpensive,
    accurate laboratory determination of the velocity of light'',
    Am. J. Phys. 41, 272--275 (1973).
  \bibitem{Crandall82}R. E. Crandall, ``Minimal apparatus for the
    speed of light measurement'', Am. J. Phys. 50(12), 1157--1159
    (1982).
  \bibitem{Becc87}F. D. Becchetti, K. C. Harvey, B. J. Schwartz, and
    M. L. Shapiro, ``Time of flight measurement of the speed of light
    using a laser and a low-voltage Pockels-cell modulator'', Am. J.
    Phys. 55(7), 632--634 (1984).
  \bibitem{Deb91}J. A. Deblaquiere, K. C. Harvey, and A. K. Hemann,
    ``Time of flight measurement of the speed of light using an
    acousto-optico modulator'', Am. J. Phys. 59 (5), 443--447 (1991).
  \bibitem{James99} Mary B. James, Robert B. Ormond and Aric
    J. Stasch, ``Speed of light measurement for the myriad'',
    Am. J. Phys. 67 (8), 681 (1999).
  \bibitem{Sakurai}\url{http://www.nep.chubu.ac.jp/v17pdf/v17-73.pdf}
    (in Japanese)
  \bibitem{tohoku}
    \url{http://www.laser.phys.tohoku.ac.jp/~yoshi/kousoku.html} (in
    Japanese).
  \bibitem{shimadzu} Shimadzu Corporation experimental equipment
    (LV--3). 
  \bibitem{HS04} Roy Henderson, Karl Schulmeister, {\sl ``Laser
    Safety''}, Taylor and Francis, (2004)    
\end{thebibliography}
\end{document}